# Uncertainty Quantification and Software Risk Analysis for Digital Twins in the Nearly Autonomous Management and Control Systems: A Review

Linyu Lin, Han Bao, Nam Dinh

**ABSTRACT**

A nearly autonomous management and control (NAMAC) system is designed to furnish recommendations to operators for achieving particular goals based on NAMAC's knowledge base. As a critical component in a NAMAC system, digital twins (DTs) are used to extract information from the knowledge base to support decision-making in reactor control and management during all modes of plant operations. With the advancement of artificial intelligence and data-driven methods, machine learning algorithms are used to build DTs of various functions in the NAMAC system. To evaluate the uncertainty of DTs and its impacts on the reactor digital instrumentation and control systems, uncertainty quantification (UQ) and software risk analysis is needed. As a comprehensive overview of prior research and a starting point for new investigations, this study selects and reviews relevant UQ techniques and software hazard and software risk analysis methods that may be suitable for DTs in the NAMAC system.

# ACRONYM

| | |
|---|---|
| ANN | Artificial Neural Network |
| BNN | Bayesian Neural Network |
| CCF | Common Cause Failure |
| DAP | Development and Assessment Process |
| DDM | Data-Driven Model |
| D3 | Defense-in-depth and diversity |
| DT | Digital Twin |
| DT-D | Digital Twin for Diagnosis |
| DT-P | Digital Twin for Prognosis |
| EBR-II | Experimental Breeder Reactor II |
| EMDAP | Evaluation Model Development and Assessment Process |
| FMEA | Failure Mode Effect Analysis |
| FTA | Fault Tree Analysis |
| HAZCADS | Hazard and Consequence Analysis for Digital Systems |
| I&C | Instrumentation and Control |
| LSTM | Long Short-Term Memory |
| ML | Machine Learning |
| NAMAC | Nearly Autonomous Management and Control |
| NPP | Nuclear Power Plant |
| PDE | Partial Differential Equation |
| PRA | Probabilistic Risk Assessment |
| RADIC | Risk Assessment for Digital I&C |
| RAVEN | Risk Analysis Virtual Environment |
| RTS | Reactor Trip System |
| SMBO | Sequential Model-based Optimization |
| SSF | Safety-Significant Factor |
| SSC | Systems, Structures, and Components |
| STPA | System-Theoretic Process Analysis |
| UQ | Uncertainty Quantification |

# 1   INTRODUCTION

In recent years, research has been widely conducted for developing and deploying autonomous control system into new reactor designs with digital twin (DT) technology, which is defined as a digital representation of a physical system that relies on real-time and history data to evaluate its complete states, to predict future behaviors, and to recommend effective actions [1]. Cetiner *et al.* [2] developed a supervisory control system that integrates probabilistic decision-making analyses and a deterministic assessment of plant state variables to support autonomous operation and control of advanced small modular reactors. Groth *et al.* [3] developed a safely managing accidental reactor transients system with dynamic probabilistic risk assessment (PRA) and counterfactual reasoning for a sodium fast reactor. All these systems use mechanistic and partial-differential-equation (PDE)-based models as DTs for inferring the safety-significant factors (SSFs) and predicting the transient of reactor states. Additionally, with the advancement of machine learning (ML) algorithms, data-driven models (DDMs) have also been used to develop DTs by learning directly from operation or simulation data and to support autonomous control by responding to the real-time measurements. Na *et al.* [4] developed an autonomous control system for a space reactor system with the model predictive control algorithm. Lee *et al.* [5] developed the autonomous transportable on-demand reactor module for preventing core damage during a loss-of-coolant accident in pressurized water reactors. DTs are developed using ML algorithms for describing the transients of reactor states at different functional hierarchies. Lin *et al.* [1] developed a nearly autonomous management and control (NAMAC) system for restoring core flow during a partial loss-of-flow accident (LOFA) in the Experimental Breeder Reactor II (EBR-II) located at Idaho National Laboratory (INL). A class of DTs is developed with different algorithms for inferring the SSFs, finding available control actions, predicting consequences, and identifying the most preferred actions [6].

When compared to previously reviewed intelligent systems, NAMAC aims to achieve an alignment of nuclear power plant (NPP) safety design, evaluation, operator training, and emergency management by extracting useful information from the knowledge base. The knowledge extraction and storage for different intended uses are achieved by various ML algorithms in DTs. The development of an ML-based DT for an intended use relies on the construction of a knowledge base and the use of advanced data-science techniques, while its deployment and operation are realized based on digital instrumentation and control (I&C) systems that consist of different software, data networks, digital platforms, and hardware carriers. Although the potential and feasibility for DTs by ML algorithms to improve safety and efficiency in reactor control have been recognized, one of the concerns for the Regulatory Commission and the nuclear industry is whether the information from a DT is credible to support decisions of high consequence and how this information could affect further DT development in an autonomous control system. To provide evidence toward the credibility assessment for autonomous control systems with DTs, uncertainty quantification (UQ) methods and software hazard and reliability analysis are needed.

The following sections are organized as shown in Figure 1. Section 2 first reviews the definition and characteristics of DT technology. Next, the DTs in a NAMAC system is discussed. Section 3 discusses the role of UQ methods and software hazard and reliability analysis in the development and assessment process (DAP) of DTs in NAMAC. Next, Section 4 reviews some specific techniques in UQ and software hazard and reliability analysis. Finally, discussions are made in Section 5 with respect to the applicability of these techniques to the ML-based DTs in the control systems.

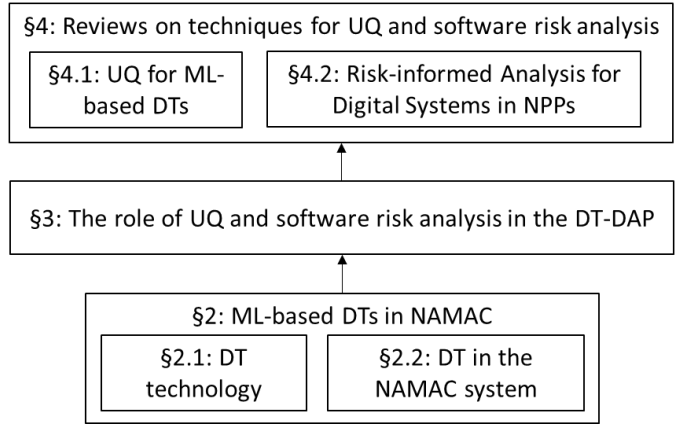

Figure 1. Overall paper structure.

## 2  MACHINE-LEARNING-BASED DIGITAL TWINS IN NAMAC

### 2.1  Digital Twin Technology

A DT is a digital representation of a physical asset or system that relies on real-time and history data for inferring complete reactor states, finding available control actions, predicting future transients, and identifying the most preferred actions [6]. During NPP operation, DTs can help decision-makers maintain an accurate understanding of the physical system, improve the effectiveness of operation, and avoid operator errors. To guide the development of DTs, the DT in NAMAC is defined based on three requirements—the DT function, the DT modeling, and the DT interface.

The DT function refers to the intended uses of a DT. In NAMAC, there are two major types of DT functions—diagnosis and prognosis. The DT for prognosis (DT-P) is used to anticipate future behaviors, while predictive performance is based on the history and the current state. The DT for diagnosis (DT-D) interrogates the current and past histories of certain objects or systems. Meanwhile, there are additional components in NAMAC like strategy inventory, prognosis, and strategy assessment, where more details can be found in [1]. This review focuses on the DTs for diagnosis and prognosis that are implemented by ML algorithms.

The DT modeling refers to techniques and algorithms that describe a virtual representation that duplicates or twins the physical system. Generally, there are four classes of options—model-free methods, model-based methods, hybrid methods, and reasoning-based methods. In the model-based approach, the model equations and parameters are derived from knowledge, including system phenomena, inner correlations, conservation laws, PDEs, etc. The resulting models are known as mechanistic, and their structure and parameters reflect some, but not necessarily all, mechanisms in the system [7]. For model-free methods, also known as data-driven methods, a generic mathematical model is fitted to the input-output pairs from the databases without explicit knowledge of phenomena, correlations, and even conservation laws. In real-world applications, the knowledge form model-based methods and correlations from model-free methods are combined to improve the accuracy and robustness of DT predictions. Such an approach is known as hybrid methods, and a classification based on different degrees of knowledge involvement in data-driven methods can be found in [8]. Moreover, reasoning-based models have recently been used for DT implementation [9], especially for the diagnosis purposes. In the NAMAC system, DTs are mostly developed by model-free methods, i.e., artificial neural network (ANN).

At last, the interface refers to the DT information transmission that is directly related to human operator, a part of the physical system. In real-time operations, it is not required for a DT to recover every detail of a physical system. Meanwhile, the required inputs for one specific system or scenario may not be applicable to other systems and scenarios. Therefore, the interface must consider the specific input/output pair for the intended uses. There could be various interface with the same function and model. For example, the diagnosis by DDMs in nominal conditions may have different input/output requirements than it would under accident scenarios because the available measurements would be different. The prognosis by system code could also have different interface requirements in steady-state calculations as opposed to transient calculations because the number of initial and boundary conditions would be different. In addition to the type of system and scenario, the interface also depends on the application program interface or operational workflow of the autonomous control system. For example, predictions by DT-P rely on an accurate diagnosis of the current reactor states in a NAMAC system [10], together with the identification of available control actions from the strategy inventory. Therefore, the input of DT-P in NAMAC must be compatible with the output of DT-D and strategy inventory.

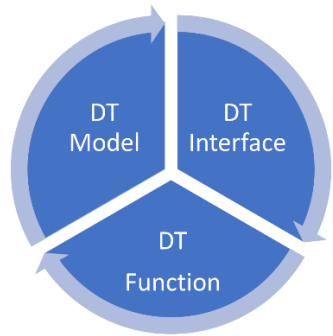

Figure 2. Definition for DTs based on three requirements.

### 2.2 Digital Twins in the Nearly Autonomous Management and Control System

To adapt to new challenges in advanced reactors, DT technology is identified as a critical component of the modern autonomous control system. In the NAMAC system by Lin *et al.* [1], an operational workflow, as shown in Figure 3, is constructed based on probabilistic risk analysis report, operating procedures, and technical specifications of nuclear reactors. A hub of DTs, including diagnosis, prognosis, strategy inventory, and strategy assessment, are implemented and connected according to the operational workflow. The NAMAC system architecture discussed here is highly modular with a hub of DTs. It is believed that this has important advantages in scalability and interpretability, which will become more important when a much broader issue space is considered. Instead of having a super DT that covers all scenarios, NAMAC aims to have a hub of DTs with different functions that combines heterogenous knowledge and that interacts with each other and with the environment. This is because in the complex system like nuclear reactor, learning process is rarely one-time training, but rather a training through different disciplines and variety of exercises. By dividing a complex learning task into a list of subtasks, each DT could perform, interact, and learn from their respective and less difficult tasks. Meanwhile, the big-data requirements on the full-scale and multi-physics system are relieved to separate-scale and simplified systems, where data is more feasible. Among all DTs, diagnosis and prognosis are critical in recovering full-flow conditions and preventing severe consequences. During accident scenarios, the DT-D reads the measurements from the reactor or simulator and produces the unobservable safety significant factors, like peak fuel and peak cladding temperature. Based on the diagnosed information, the strategy inventory is to identify all available control actions for mitigating the situation or shutting down the reactor. Meanwhile, the DT-P is activated for doing front running predictions that evaluate the consequences of different actions from the inventory and the initiating events from the DT-D. At last, these predictions are fed into the strategy assessment for finding the optimal path that satisfies specific goals and preference. At the same time, to stress the importance of AI trustworthiness, a discrepancy checker is included to monitor the differences between NAMAC expectations and observations. Measured by a distance metric, if the differences are larger than a limit, the DTs' outputs or NAMAC's recommendations can no longer be trusted either because it is operated outside the domain or there is too much uncertainty. In this case, a safety-minded signal, for example, scram command, will be send to the operator.

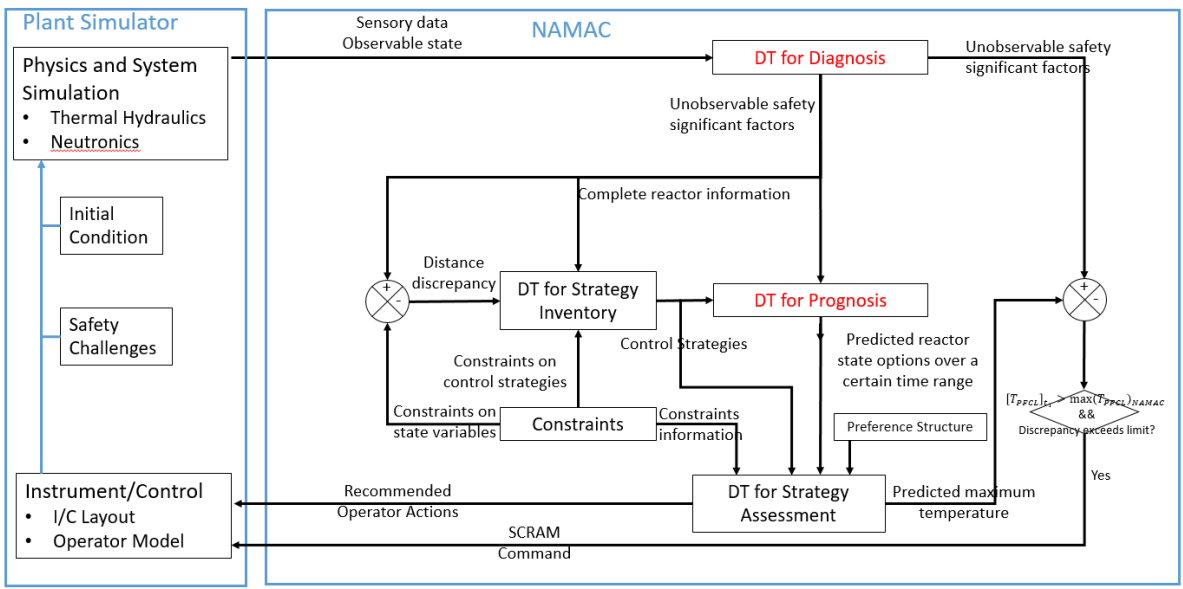

Figure 3. NAMAC operational workflow. DTs of different functions are connected and further coupled with the plant simulator for reading sensor data and injecting recommendations.

In the NAMAC system, machine learning algorithms, especially the artificial neural networks (ANNs), are used to implement DTs. The goal is to be adaptable and scalable with respect to issue space and to be friendly to big data. Although ML promises to find rules that are probably correct about most members of the set they concern [11], the no-free-lunch theory for ML states that when averaged over all possible data-generating distributions, every classification algorithm has the same error rate when classifying previously unobserved points [12]. In other words, no ML algorithm and learning scheme is universally better than any other. The best algorithm has the same average performance as merely predicting that every point belongs to the same class. As a result, a DAP is needed for DTs to learn and optimize the training and application of ML algorithms [1] [13]. The training process includes the tuning of hyperparameters, neural network architecture, learning parameters, etc., while the application includes the integration of ML-based DTs into a NAMAC system, the deployment of NAMAC into a reactor I&C system, the use of NAMAC system during accident scenarios, etc.

# 3   THE ROLE OF UQ AND SOFTWARE RISK ANALYSIS IN THE DT-DAP

The DAP is defined as a formalized process to ensure applications are transparently and consistently developed, assessed, and decided in accordance with the target expectations as set out in the planning stage. To furnish a recommendation, the NAMAC system is characterized by an operational workflow with interacting and reacting DTs consisting of different knowledge base sets/subsets. The development and performance of interacting DTs vary from one operation condition to another. To implement such a system, the details of each DT are analyzed separately from a reductionist point of view. However, in view of DT interactions and associated synergistic effects, such an approach may not lead to a credible NAMAC recommendation. Such an issue becomes critical when the implemented system is applied to different operational conditions. Since the generalization capability depend on the trade-off between data-distribution and network complexity [14], when DTs of different generalization capabilities are connected by the operational workflow, the credibility of NAMAC recommendation can hardly be traceable with only separate assessments. As a result, global considerations are needed to evaluate the interactive and synergistic effects such that the organization among complex interactions can be established. In many established DAPs and standards, a two-tiered approach, one based on the top-down or system approach, the other on the bottom-up or separate approach, is predicted by:

1. The need to minimize arbitrariness in deriving acceptance criteria.
2. The need to develop a methodology that is transparent, consistent, and practical.
3. The need to provide a technical justifiable basis for deciding the adequacy of DTs and NAMAC for the target applications in a timely and cost-effective manner.

The top-down approach is used to evaluate the impact of NAMAC software to the reactor digital I&C system and identify important DTs that need to be developed and assessed in greater detail by conducting a bottom-up analysis. The bottom-up approach focuses on the development and assessment of separate and important DTs. It conducts a detailed analysis such that all important requirements are fully addressed, and thereby, ensures the performance of DTs in the target applications.

Based on this principle, a Digital Twin Development and Assessment Process (DT-DAP) was recently proposed for the NAMAC system based on the Evaluation Model Development and Assessment Process (EMDAP) [3]. EMDAP is a framework that integrates requirements, knowledge-based generation, model development, and assessment throughout the life cycle of models. Developed by the U.S. Nuclear Regulatory Commission (NRC), EMDAP aims to describe an acceptable process for developing and assessing the evaluation models used to analyze transient and accident behaviors within the design basis of an NPP. The principle of developing and applying an EMDAP in a study on quantifying reactor safety margins (NUREG/CR5249, Ref. 3), which applies the code scaling, applicability, and uncertainty evaluation methodology to a large-break loss of coolant accident. The purpose of that study is to demonstrate a method that could be used to quantify uncertainties as required by the best-estimate option described in the NRC's 1988 revision to the Emergency Core Cooling System (ECCS) Rule (10 CFR 50.46) [15]. In EMDAP, the evaluation model assessment is divided into two parts. The first pertains to the bottom-up evaluation of the closure relations for each code, including their pedigree, applicability, fidelity to appropriate fundamental or separate effect data, and scalability. The second pertains to the top-down evaluations of code-governing equations, numeric, the integrated performance of each code, and the integrated performance of the overall evaluation model. Calculations of actual plant transients or large-scale integral effect tests are used as confirmatory supporting assessment for the top-down evaluations. Compared to the first part, the second usually has less resolution to determine the adequacy of individual models. However, it focuses on the applicability of EM to the target system and transient and evaluates the scalability of integrated calculations and data for distortions.

On one hand, it is critical to evaluate the adequacy of DTs based on similar principles in the EMDAP framework. On the other hand, since the DT is to support real-time reactor operations within the NAMAC system, it is deployed as an active component of control software in the reactor digital I&C system. Considering the significance of an I&C system to the safety of reactor operations, software hazard and reliability analysis is needed as a top-down evaluation for DTs. As a result, six basic principles are suggested in the DT-DAP:

(1) **Establish requirements for DT and NAMAC system capability**. The purpose of this element is to establish requirements for the DT, including the purpose of the control system, transient class, interface with the DT, model, and function, reactor types, etc.

(2) **Develop knowledge base.** Based on requirements, this element aims to construct a knowledge base including the issue space, simulation tool, and data repository. The issue space defines the scenario in mathematical formulations. The simulation tool is required to generate the training and testing data for the development and assessment of different DTs. The data repository has two elements: the knowledge element and the data element. The knowledge element includes reports or documents related to the safety design, analysis, operation, training, and benchmarking, etc. The data element consists of data generated by simulation tools and testing facilities for the development of DTs [16].

(3) **Develop DTs**. Element 3 develops DTs based on the knowledge base generated by Element 2 and the requirements by Element 1. For ML-base DTs, Element 3 establishes a training plan, including the training scheme, sets of hyperparameters, input/output features, normalization techniques, etc.

(4) **Assess the uncertainty of DT and NAMAC.** Based on the knowledge base, the DTs are assessed in two parts: a bottom-up approach and a top-down approach. The bottom-up approach evaluates the uncertainty of separate DTs, including their accuracy, generalization capability, and fidelity to proper subsets of the knowledge base. The top-down approach is to analyze the overall vulnerabilities, reliability, and impact of NAMAC as a software in the reactor I&C system. An integrated risk-informed analysis including software hazard analysis, reliability analysis, and consequence analysis should be performed to provide sufficient best-estimate results for system designers, which also provides technical support for detecting, preventing, and mitigating different types of cyber threats.

(5) **Assess adequacy.** The objective of this element is to evaluate the adequacy of DTs and NAMAC system based on the requirements, knowledge base, DT implementation, and DT assessment results. If DTs and the corresponding system are adequate, they are applied to target reactors, transients, and scenarios for risk analysis. If DTs and NAMAC do not meet the adequacy standard, users need to return to appropriate elements for corrections. The iteration will continue until all adequacy standards are met. For complex systems and scenarios, the assessment and decision-making should be conducted in a transparent, consistent, and improvable manner. Peer reviews by independent experts, quantitative frameworks for adequacy assessment, software requirements, etc., should be an integral part of the adequacy assessment process.

(6) **Provide comprehensive, accurate, and up-to-date documentation.** This element is designed for clear and credible adequacy assessment and decision-making. Since the DAP could lead to changes in importance determination, it is important that documentation of this activity be developed early and kept current.

Based on these principles, Figure 4 adapts the EMDAP to the DT-DAP for implementing a credible NAMAC with a set of acceptable DTs for furnishing recommendations to operators. The first five principles correspond to Element 1 to 5, respectively. Considering the complexity of the reactor system and the development process, the DT-DAP can be classified into two phases: scoping and refinement. In all phases, it is required that DT function and interface be implemented with the designed modeling approach for the target applications. Meanwhile, the DT errors are required to be evaluated such that they satisfy the acceptance criteria. At the scoping stage, the DAP is mainly driven by expert knowledge and user experiences while assessment techniques are largely qualitative, including both sensitivity and uncertainty analysis. The goal is to identify dominant sources of uncertainty, evaluate their relative importance, and optimize the training plan based on experiences. However, there can be large uncertainty and bias due to the limited knowledge of user groups. Although the DT performance satisfies the acceptance criteria, it could be only applicable in the training domain or a specific class of scenarios. As a result, when the DTs and control system are applied to different transients, their uncertainty can be so large that it could alter other DTs and final decisions. In such conditions, the adequacy of the DT and the control system needs to be carefully investigated.

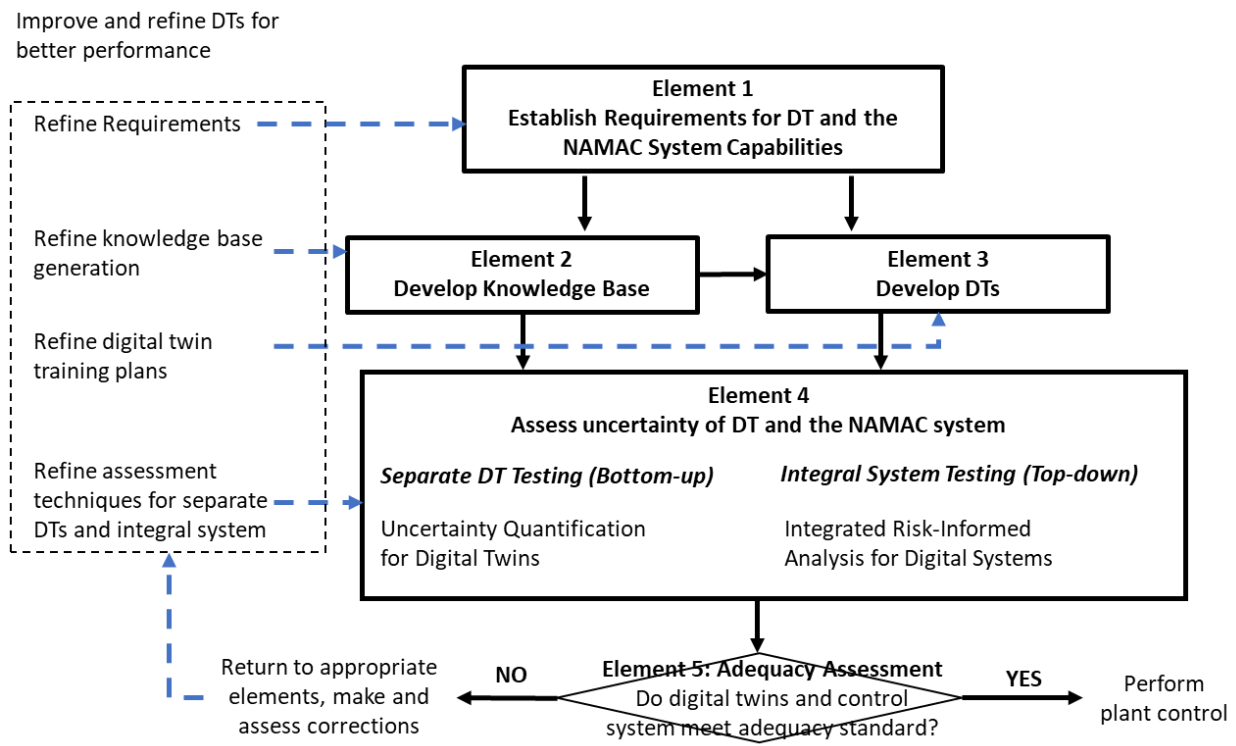

Figure 4. Scheme for the DT-DAP. Adopted from the EMDAP framework [3].

In a NAMAC case study [1], DT-D and DT-P are implemented with ML algorithms (i.e., feedforward neural nets) and assessed as separate DTs within the NAMAC system. Requirements (Element 1) are set up based on the NAMAC objective and operational workflow. Meanwhile, EBR-II and a partial LOFA due to primary sodium pump (PSP) malfunctions are selected as the type of reactor and transient, respectively. A knowledge base (Element 2) is constructed by first characterizing the LOFA issue space based on the speed curves of two PSPs. Next, a GOTHIC™ model for the primary side of EBR-II is constructed and benchmarked against experimental data [17]. A data repository is built by coupling GOTHIC with a statistical software named the Risk Analysis Virtual Environment (RAVEN) [18] [19] for sampling, data generation, and storage. To train ML algorithms for DT-D and DT-P (Element 3), the database is first separated into training and testing according to the issue space characterizations. To test the generalization capability of DTs, the testing data contains "unseen" data points outside the training domain. Driven by the ML errors and user experience in ML training, the ML-based DT-D and DT-P are developed and improved. At last, the training root mean square errors equal 1.42℃ and 0.54℃, respectively. To assess the performance of the DTs, they are first evaluated separately. It is found that the L1 relative error norms of both DT-D and DT-P in the testing domain are less than 5%. Other sources of uncertainty, including input time ranges, sensor failures, and target training losses, are evaluated and found to be strongly correlated with the error of DTs. Next, these DTs are assembled into NAMAC, and the NAMAC is further tested by coupling it with the GOTHIC plant simulator. A confusion matrix [20] is defined to quantify the accuracy of the NAMAC system in classifying the predicted consequence of control actions. It is found, when NAMAC is operated within the training issue space, the confusion ratio equals zero, indicating a reasonable recommendation that is consistent with the historical norm of human operation. However, when the issue space is outside the training domain, the confusion ratio grows, indicating the recommendations are either too conservative or could lead to component failure. To further reduce the uncertainty of NAMAC, advanced ML algorithms, different network architectures, samples of hyperparameters, etc., are used.

It can be found from this DT-DAP that UQ methods play key roles in the bottom-up approach for evaluating the performance of separate DTs with respect to different sources of uncertainty. The software risk analysis is critical to the top-down approach, which determines the potential failures of the NAMAC system, impacts to the reactor I&C systems, reliability of control systems and digital SSCs, and safety consequences during reactor operations. To guide the selection and further development of specific techniques, the following sections review a class of UQ and software risk analysis. Section 4.1 reviews the definition for UQ together with three classical UQ methods and one newly proposed meta-learning for ANN-based DTs. Section 4.2 reviews methods for software hazard, reliability, and consequence analyses, together with an integrated risk assessment strategy for digital I&C systems.

# 4 REVIEWS ON TECHNIQUES FOR UNCERTAINTY QUANTIFICATION AND SOFTWARE RISK ANALYSIS

## 4.1 Uncertainty Quantification for Machine-Learning-based Digital Twins

As a critical technique to the separate DT assessment, various UQ techniques can be applied to identify the sources of uncertainty and evaluate their impacts on their DT outputs. Since NAMAC adopts ANNs, classical techniques—including Bayesian inference—interval analysis, and fuzzy theory, are reviewed first. They are classified as classical since these methods generally follow the flowchart in Figure 5. At the same time, a newly proposed meta-learning technique is reviewed based on an updated scheme.

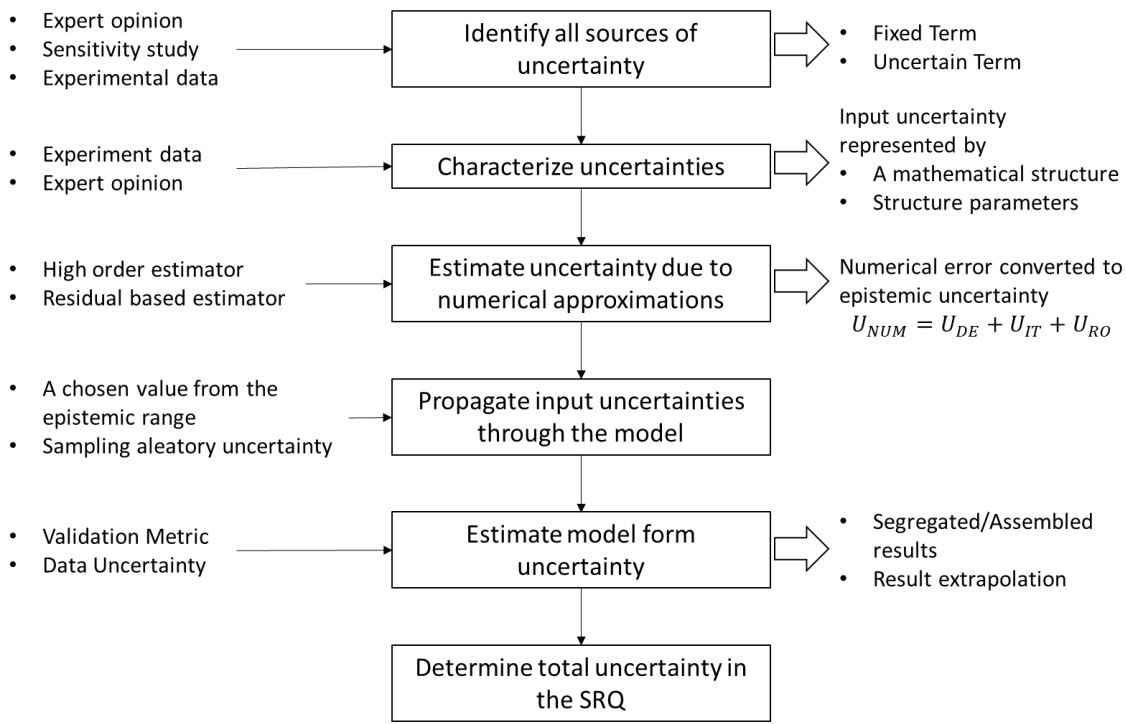

Figure 5. Schematic flowchart of uncertainty analysis, including examples of inputs and outputs for each component.

### 4.1.1. Uncertainty Quantification Methods

In the process of verification and validation, UQ is an inherent procedure that refers to the activity of identifying and understanding all possible uncertainties within the system of interest. From the fundamental essence, uncertainty can be classified into two categories: epistemic and aleatory uncertainty [21]. The aleatory uncertainty is induced by inherent variation or randomness. Although aleatory uncertainty is not reducible, it can be accurately described with a sufficiently large number of samples. Epistemic uncertainty arises due to a lack of knowledge, and if sufficient knowledge is added, the epistemic uncertainty can be eliminated [22]. In scientific applications, it is usually hard to distinguish these two uncertainties. Depending on the questions, uncertainty classifications can be different for the same problem. It is also argued in some literature [21] that the aleatory can be interpreted as epistemic uncertainty by the principle of direct inference. In this study, discussions will be limited to the treatment of epistemic uncertainty. Two approaches exist for representing an epistemic uncertainty: the interval with no associated probability distribution function (PDF) or the degree of belief by a PDF. The latter approach is usually known as the Bayesian approach to UQ while the former is usually based on the interval arithmetic, where no value is truer than any other value. Detailed discussion is beyond the scope of this study, which mainly investigates the Bayesian approach.

In scientific computing, it is important to identify all sources of uncertainty. If a fixed value is known precisely, the simulation uncertainties can be treated as deterministic. Otherwise, they are represented with probability. There are three sources of uncertainty: model inputs, model form, and numerical approximation. Model inputs mainly include the model parameters, initial and boundary conditions, and geometry. Usually, the input uncertainties are characterized or calibrated based on expert opinion, measurements, theories, etc. If the input uncertainties are represented by a distribution function, they will be propagated by performing a number of individual simulations with sampled inputs. The number of simulations depends on the problem specifications, including the linearity of the equation systems, dependence among different inputs, sampling techniques, etc. The model form uncertainty mainly comes from the assumptions that models rely on [23], and it is usually characterized by the validation process. For models that are built on PDEs, numerical approximations become an important source of uncertainty. The process of determining uncertainties associated with numerical approximations is known as verification. Specifically, the uncertainties of numerical approximations can be further divided into four components: discretization error, iterative convergence error, round-off error, and programming mistakes. Various methods have been developed for estimating numerical error, including a higher-order estimator [24] [25] and residual-based estimator [26]. Engineering standards like the grid-convergence index method have also been made to guide mesh convergence studies [27] [28]. Figure 5 previously showed a schematic flowchart for the UQ process. Examples of inputs and outputs for each component are shown in the figure.

The UQ framework provides a consistent and straightforward scheme for estimating and propagating the uncertainty to the total uncertainty. This study reviews three algorithms for characterizing and propagating uncertainty, including Bayesian inference, interval analysis, and fuzzy sets. Bayesian inference and interval analysis both describe uncertainty by probability theory, and the probability is treated as a degree of belief in a hypothesis. Bayesian inference assumes a prior distribution can be quantified or elicited precisely, and there is no distinction between epistemic and aleatory uncertainty. Interval analysis assumes that the true value locates somewhere within the range, and this fact does not imply a uniform distribution over this interval. Meanwhile, Ferson *et al*. [29] argue there are two kinds of uncertainty—one arises as a variability resulting from heterogeneity or stochasticity while the other arises as partial ignorance resulting from systematic measurement error or subjective uncertainty. As a result, the interval analysis should be used to propagate ignorance, and probability theory is used to propagate variability. The fuzzy set is an extension of the interval analysis, but with gradual assignment within the range:

- Bayesian Inference: Within the Bayesian inference framework [30], a prior distribution $f_X(x)$ for the uncertain parameters of the model is first assumed. Next, such a distribution encapsulates the current understanding of the values and provides a likelihood function, $P(B|X = x)$. Using this prior distribution and the results of the simulation or experiment $P(B)$, the posterior distribution $f(x|B)$ can be generated via the Bayes formula: $f(x|B) = P(B|X = x)f_X(x)/P(B)$. With Bayesian inference, the prior statistical analysis of a model parameter can be updated based on additional simulation or experiment results. The updated statistical properties for the uncertain parameter are characterized by the posterior distribution. These properties can be directly sampled from the distribution [31] and propagated through the simulation. To support the development of ANNs, Bayesian inference is used by assigning prior uncertainty on network weights; thus, it provides uncertainty about the functional mean through posterior predictive distributions [32]. These networks are known as Bayesian neural networks (BNNs). In literature, posteriors inferred by Hamiltonian Monte Carlo [33] are frequently used as ground truth. Since Hamiltonian Monte Carlo scales poorly on high-dimensional parameter space and large datasets [34], efforts have been put on variational methods—including Bayes by backprop [35], matrix-variate Gaussian [36], Bayes by hypernet [37], etc. However, Yao *et al.* found that the capability of BNN in capturing application data depends on the selection of inference methods and error metrics [38]. Meanwhile, it is found that BNN shows limited capability in predicting data outside the training domain as compared to non-Bayesian methods.

- Interval Analysis: Interval analysis [39] is developed to represent uncertainty with imprecise probability. Assume $x_i \in X$ is one of the inputs to the model $y = F(X, \{a\})$ and $x_i$ belongs to a certain interval $x_i \subset [a, b]$. The interval analysis aims to determine the minimum and maximum values of $y$ with either naïve [40] or optimization approach. For naïve approaches, uncertainty is propagated with interval arithmetic: Every variable is represented as two numbers, which bound the true value of the variable. Therefore, the uncertainty propagation is made by simply changing the type of variables in the computer code from a single-number type to an interval. Although this extension to interval variables can be useful in some instances, the so-called dependency problem reduces the uses of such methods in general [40] and tends to overestimate the final uncertainty. Muhanna *et al*. [40] have processed the interval quantities through finite element computations, where the uncertainty overestimation is almost eliminated. However, such techniques have only been applied to linear problems, and their capability in complex system remains unknown. To deal with functions with large input spaces, optimization techniques, like gradient descent methods or evolutionary optimization schemes, are used to search for the minimum and maximum values of $y$. Interval analysis [39] has also been applied to quantify uncertainty and to optimize neural network outputs [41]. However, for a complex system where different sources of uncertainty are not independent, the bounds can be overestimated by interval analysis [40]. Indeed, the drawbacks of interval analysis—i.e. dependency effect and high computational load—can hardly be avoided. Although with Taylor models, the interval analysis results in an excellent convergence for small intervals, the dependency effect remains and becomes even worse when evaluating large input domains [42]. Meanwhile, because interval analysis treats information in terms of binaries, a gradual assignment of elements to the interval or a weighting of elements within the interval cannot be accounted for. Consequently, a degree of confidence that a particular event occurs cannot be deduced with the aid of interval quantities alone.

- Fuzzy Theory [43]: As an extension of the interval analysis, the interval of fuzzy sets is assessed or weighted with the aid of membership values from a continuous function. There are three main semantics for membership function: similarity, preference, and uncertainty [44]. For the uncertainty semantic, membership values can be treated as a subjective assessment based on the degree of epistemic possibility. Meanwhile, expert evaluations and imprecise values can be integrated into the fuzzy set, including linguistic variables, uncertain measurements, etc. More details about fuzzy modeling based on imprecise data are discussed in [45]. Because fuzzy systems are able to approximate any nonlinear function within the range of a desired accuracy, fuzzy logics are combined with the neural networks, known as a neuro-fuzzy system (NFS) for modeling and control [46] [47]. Pezeshki *et al.* have compared NFS against ANN to predict building thermal consumption outside the training domain [48]. Although NFS has better generalization capability, the computations are more time-consuming, and learning is also expensive. This is caused by the complex underlying deterministic computation (mapping model). When the number of rules gets larger, the model delays become high and can hardly be used for problems of industrial size.

### 4.1.2. Meta-Learning

When people learn new skills, they usually start from skills learned earlier in related tasks, reuse approaches that worked well before, and focus on what is likely worth trying based on experience [49]. Similarly, when ML models are trained for a specific task, experiences with related training tasks could also improve the training with fewer iterations and errors. These techniques are known as meta-learning, and it divides one big learning task, which is expected to fit to a range of new tasks, into two hierarchies of learnings: base learner and meta learner. Compared to the classical training for ML-based DTs, denoted as the base learner, a meta model exists that optimizes the base learner by updating its parameters via a meta knowledge base. As shown in Figure 6, assume there is a set of parameters, $\theta$, which directly affect base learner performance and is highly adaptable to task properties and applications. During the course of meta-learning, denoted by the bold line, meta learner $L$ for a set of task $[t_1, t_2, t_3 \ldots]$ optimizes parameter $\theta$ such that when a gradient step is taken with respect to a particular task, $t_i$, denoted by the thin lines, the parameters are close to the optimal parameters $\theta_i^*$ for task $t_i$. Meta-learning techniques do not make assumptions on the form of the model, and the parameters for meta-learner, $L$, known as meta-features, are extracted from the knowledge base.

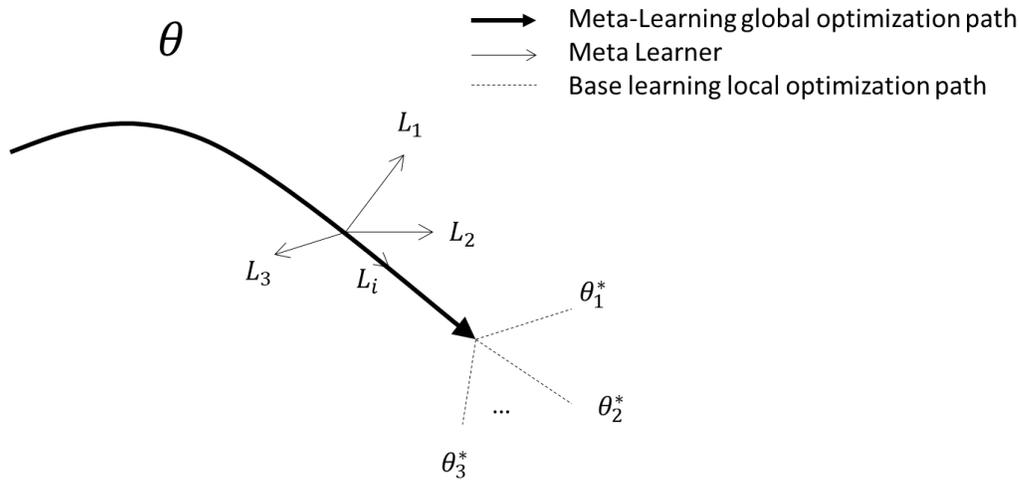

Figure 6. Diagram of meta-learning techniques.

Figure 7 shows the flowchart for meta-learning techniques aiming to find the best ML-based DTs for new tasks based on experiences from prior tasks. The flowchart is adapted from the general learning scheme for meta-learning by Brazdil *et al*. [50]. First, given a knowledge base from prior tasks, a meta-knowledge base is constructed by extracting meta-features from the knowledge base. Next, for a new task, meta-features are extracted and compared with the prior task meta knowledge base for searching, evaluating, and matching an optimized ML model that could maximally reach the acceptance criteria. Red blocks with round corners represent two major steps of meta-learning [51]:

(a) A meta knowledge base describes prior learning tasks, evaluation results, and previously learned models. Depending on meta-feature extraction techniques, prior learning tasks can be: (1) the exact algorithm configurations used to train the models, like hyperparameter settings, pipeline compositions and/or network architectures; (2) the task itself, like the statistical properties of databases, issue spaces, and differences in initial/boundary conditions in the simulation tools; and (3) the evaluation results, like model accuracy, model performance, or training time.

(b) Based on the prior meta knowledge base, new task properties, and the meta-feature extraction techniques, meta-learners are developed to search, match, or explore the best ML models. For evaluation-based techniques, the best model is explored by searching for the optimal ML configurations from the meta knowledge base that could maximize the performance function. For task-based techniques, prior tasks that best match new tasks are identified, and ML models with similar configurations to the prior tasks are selected. For prior-model-based techniques, an initial guess is made with the prior model. New models are explored based on prior knowledge and corresponding feedback.

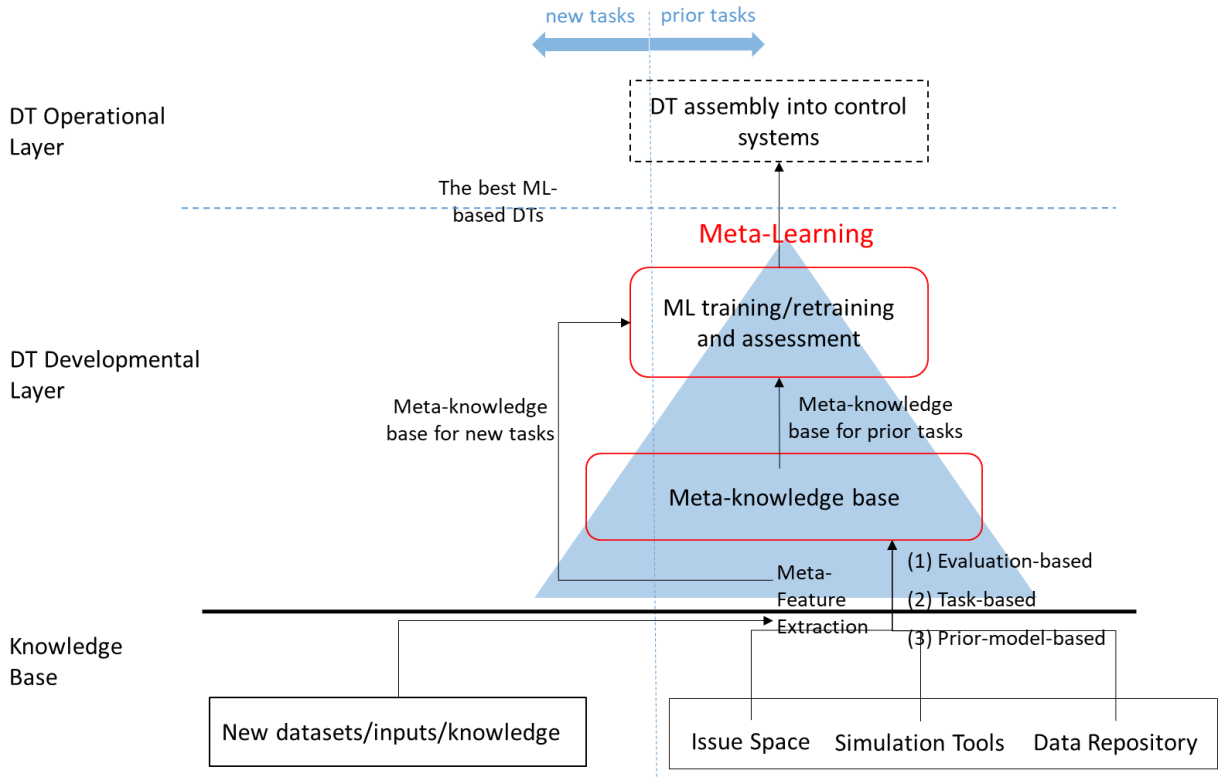

Figure 7. Schematic flowchart for meta-learning algorithms that aims to find the best ML algorithm and model.

These two components make up meta-learning, and such a task is used to be performed by a single and big ML training and assessment process. After the implementation of meta-learning, the best ML-based DT is assembled in the autonomous control system or reactor system for the intended uses. It is stressed that either the meta knowledge base or the ML training and retraining cannot deviate completely from the prior tasks, even though there are new meta knowledge bases extracted from new datasets, inputs, and knowledge. Based on the current review on meta-learning, there still needs to be a certain amount of overlap between the new and prior tasks to ensure ML performance [50] [51]. The following sections review how these three classes of meta-feature extraction and corresponding meta-learner identify the best ML-based DTs for a new task in detail:

- Learning from Model Evaluations: Consider a set of prior tasks $t_j \in T$ and a set of learning algorithms defined by their configurations $\theta_i \in \Theta$, where configurations can be hyperparameter settings, network architecture components, or training parameters. Assuming that **P** is the set of all prior scale evaluations $P_{i,j} = P(\theta_i, t_j)$ for configuration $\theta_i$ and task $t_j$, based on predefined evaluation measures like accuracy, or model-evaluation techniques, like cross validation. $\mathbf{P}_{new}$ is the set of new evaluations $P_{i,new}$ on a new task $t_{new}$. The objective of meta-learning algorithms is to have a meta-learner $L$ such that the recommended configurations $\Theta_{new}$ can be found for a new task $t_{new}$. The meta-learner $L$ is usually trained by predefined meta-data repositories, which aims to predict $\mathbf{P}_{new}$ based on the configurations. One of the popular approaches is sequential model-based optimization (SMBO) [52]. As a model-based approach, SMBO approximates $L$ with a surrogate model, like expected improvement [53], conditional entropy of the minimizer [54], and bandit-based criterion [55].

- Learning from Task Properties: Considering there are a set of prior tasks $t_j \in T$, while each task can be described with a vector $M = (m_{j,1}, \dots, m_{j,K})$ of $K$ meta-features. These features can be used to define the task similarity and information from the most similar tasks can be transferred to the new task $t_{new}$. Moreover, a meta model $L$ can be constructed between a task's meta-features and the utility of specific configurations $\Theta_{new}$ given the meta-features $M_{new}$ of the new task $t_{new}$. In addition to some commonly used meta-features like ANalysis Of Variance (ANOVA) p-value, principal component analysis (PCA) skewness, covariance, correlation, etc. [56], a model that generates a landmark-like meta-feature representation $M'$ given other task meta-features can be trained by performance data **P**. Sun *et al.* [57] evaluated a predefined set of configurations $\theta_i$ on all prior tasks $t_j$, and generated a binary meta-feature $m_{j,a,b} \in M$ for every pairwise combination of configurations $\theta_a$ and $\theta_b$, which indicates where $\theta_a$ has better performance than $\theta_b$. As a result, a set of meta-feature representation $M' = (m_{j,a,b}, m_{j,a,c}, m_{j,b,c}, \dots)$ can be generated that compares every pairwise combination of ML configurations. Perrone *et al.* [58] have trained a feedforward neural network that learns and expands the original ML configurations into $\theta'$ such that a linear surrogate model can accurately predict $\mathbf{P}_{new}$ with $\theta'$. In [59], a concept of physics coverage condition (PCC) was defined to represent the coverage or similarity of existing data/tasks and target data/tasks by distinguishing task properties based on both global and local characteristics. The global condition includes dimension, geometry, physics, and statistical features that are determined based on domain knowledge. The local condition covers variables and information in local cells or fields and data distributions. It is found that new tasks with completely new global characteristics can be very similar to prior tasks when their local characteristics are identified [60].

- Learning from Prior Models: In previous methods, a meta-learner $L$ is trained based on the predefined meta-data or meta-features and applied to a new task $t_{new}$. However, this third type of algorithm uses ML models by prior tasks $t_j$ as starting points for creating new ML models for new tasks $t_{new}$. Such algorithms are also known as transfer-learning [61]. They have been used for optimizing ANN hyperparameters and architectures. Schmidbuher [62] developed a method that can modify the weight of neural networks based on observed errors. The updating scheme is a differentiable parametric model, and both the network and learning algorithms are trained jointly by gradient descent. Bengio *et al*. [63] proposed learning updates by parametric rules, the parameters of which are optimized with a gradient descent method. Because simple rules may not be applicable to complex tasks, Runarsson *et al*. replaced those rules with a neural network [64]. Younger *et al*. [65] connected the above transfer-learning threads by allowing for the output of backpropagation from one network to feed into an additional learning network, while both networks are trained jointly. The forms of prior models are not limited to DDMs; a meta learner can learn from physical laws, mechanistic models, and reasoning rules as well. Ioannou [66] proposes a structural prior for convolutional neural networks. A convolutional filter is learned from a collection of basis filters for reducing the spatial and channel connectivity in conventional neural networks without compromising accuracy. Seo and Liu [67] created a differentiable physics-informed graph network to incorporate implicit physics knowledge from differentiable physics equations. The graphic model handles spatially located objects and their relations from equations as vertices and edges. Meanwhile, a recurrent computation is developed to learn the temporal dependencies. Raissi *et al.* [68] combined a physics-informed and physics-uninformed neural network for coupling the dynamics of passive scalar and Navier-Stoke equations with automatic differentiation. During the training, the residuals velocity and pressure fields at a finite set of points based on the Navier-Stoke and transport equation are incorporated into the loss function. By minimizing the loss function with physics "penalty," Navier-Stokes equations are "encoded" into the neural network while being agnostic to the geometry or the initial and boundary conditions.

Classical UQ techniques can be used in the bottom-up DAP for evaluating and improving the performance of separate DTs. It is suggested from the review that the specific selection depends on the target problem and requirements. For example, Bayesian inference is suitable for inverse problems, with model parameters as the key sources of uncertainty, and it usually needs an accurate estimate on the prior distribution for uncertain parameters. Interval analysis can distinguish the different types of uncertainty, but it suffers from a dependency problem that requires high computational sources. Fuzzy theory can be combined with neural networks as NFS for improving their generalization capability. It can also be used for linguistic expressions and expert inputs. However, for complex or fast-transient problems, NFS can have large model delays and the training becomes computationally expensive. Moreover, classical UQ techniques have not considered the effect of data quality, data quantity, or domain knowledge, which all have proven to be critical to the generalization capability of ML algorithms [11]. Meta-learning is a newly proposed concept for improving the learning and generalization capability of neural networks with three major techniques. Instead of starting every UQ task from scratch or making a guess, meta-learning stresses the importance of prior knowledge. The objective is to identify and extract useful information in model evaluation results, prior task properties, and prior model for the new tasks. It is believed that as more tasks are learned and more prior knowledge is stored by the meta learner, the new task will become more similar and the generalization capability of ANNs will be naturally improved. Among all three meta-learning techniques, learning from the prior model, or transfer-learning, appears to be the most effective approach in incorporating and leveraging the underlying physics laws and governing equations to the development of ML and ANNs.

## 4.2 Risk-informed Analysis for Digital Systems in Nuclear Power Plants

For a digital-based I&C system, like an AI-guided autonomous control system, the failure of the I&C function results from either a hardware or software/digital failure which may result from the uncertainties AI/ML predictive capabilities. Recent development and deployment of digital I&C technologies in NPPs presented some technical issues, such as software common cause failures (CCFs) and the lack of onsite plant experience, especially for safety-related systems [74] [75]. Therefore, safety-related digital I&C systems in NPPs are required to have diverse and redundant designs for prevention and mitigation of specific CCFs [75]. Although activities have been made to develop digital system models (particularly for software CCFs), no suitable approaches have been generally accepted in current NPP PRA efforts considering the complexity of failure causes and system designs. For AI-guided autonomous control systems, some systematic and multiple inappropriate automated and automatic actions could be actuated because of not only traditional digital/software CCFs, but also uncertainties of AI/ML predictions or specific cyber-attacks to embedded AI/ML. Consequences of these inappropriate actions may significantly affect reactor safety, especially when these actions are relevant to safety-related or safety-significant components. In June 2020, the NRC published a new version of NUREG-0800 "Guidance for evaluation of potential common cause failure in digital I&C systems", which provided guidance for reviewing a licensee or applicant's evaluation of 1) a digital I&C system's vulnerability to CCF due to latent defects in the software or software-based logic; 2) the effects of such a CCF on plant safety; and 3) the measures implemented to limit, mitigate, or cope with the effects of the CCF. Digital I&C systems are categorized into four types according to whether they are safety-related or safety-significant. The assessment for the safety-related and safety-significant systems (defined as A1 systems) should consider the susceptibility to CCF of the integrated system and the consequences of CCFs that could affect the integrated or interconnected A1 systems. Accordingly, a defense-in-depth and diversity (D3) assessment should be performed to verify the design maintains defense-in-depth and meets applicable requirements.

For those autonomous control systems designed for safety purposes, diverse and redundant features should be designed to avoid single points of failure. Accordingly, software CCFs may occur in different DT software for different intended uses in one autonomous control system, or redundant DTs for the same intended use. Therefore, suitable methods for the hazard and reliability analysis of DT software should be selected and applied to analyze the vulnerability and reliability of DT, particularly to potential software CCFs in a NAMAC system.

### 4.2.1. Methods for Software Hazard Analysis

Software hazard analysis aims to identify the root causes that may lead to functional failure, unintended function, inadvertent function, or operational degradation of the target software. These root causes can be design defects of software, external disturbances, hardware random failures, cyber-attacks, or human errors. Hardware failure modes and human errors (of operators, engineers, designers) can be identified using traditional hazard analysis methods like failure mode effect analysis (FMEA) and fault tree analysis (FTA), which have been performed for PRA of analog systems in NPPs in past decades. However, systematic interactions between digital systems and other NPP components, between internal components of one digital system and between different digital systems that often result in some unanalyzed digital failure modes are difficult for FMES and FTA to track [69]. One potential major concern in the licensing of new DT software designs for autonomous controls is the uncertain risks resulting from CCFs in DT software, particularly in the control systems with safety features that have redundant designs. It has been observed in NRC staff reviews of failure modes in digital safety systems that FMEA does not address CCFs that result from systemic causes such as engineering deficiencies. [70] In addition to systematic interactions, there are some other factors determining these traditional successful hazard analysis methods for analog systems not suitable for analyzing software or digital failure modes in advanced digital AI-guided systems. First, technically, software or digital platforms never "fail" because they are designed or programmed by human for some specific functions or conditions and always perform as what are designed inside. However, some trivial assumptions made or restrict requirements followed during software development life cycle (SDLC) may not meet some extreme conditions or unexpected conditions in new operating environment due to upgrades. It is shown that a majority of software issues causing serious accidents have been investigated as design defects or incompleteness of requirements in SDLC, other than in the implementation process of these requirements [84].

To investigate how undesired systematic failures can result from inappropriate assumptions or incompleteness of design requirements during SDLC, a hazard analysis method called systems-theoretic process analysis (STPA) was developed by Leveson [71] to identify systematic hazards that may lead to undesired or unexpected losses. STPA was performed to investigate the risk of a digital main steam-isolation valve in NPPs in 2012. [72] In 2014, STPA was also performed for the hazard analysis of an aircraft automatic braking. [73] However, as a qualitative hazard analysis method, STPA only focuses on identifying systematic hazards and providing recommendations on elimination of causal factors, it doesn't provide quantitative risk-informed information for the estimation of hazard consequences, which is important for the D3 assessment of safety-related and safety-significant digital I&C systems in NPPs. Then hazard and consequence analysis for digital systems (HAZCADS), a novel approach for hazard analysis was developed by the Electric Power Research Institute and Sandia National Laboratories to build an integrated fault tree by adding applicable software failures identified by STPA as basic events into the hardware fault tree built by FTA. [74] STPA and HAZACADS are both recent advancement in hazard analysis of digital systems, however, technical approaches were not developed to identify and quantify potential software/digital CCFs that may become significant due to the complexity of redundant designs in both traditional safety-related digital systems and AI-guided autonomous control systems with multiple automated safety features. Therefore, a new hazard analysis method of DT software should be developed to systematically analyze the vulnerability of DT software and provide technical basis for DT software reliability analysis, particularly to potential software CCFs that may be caused by AI/ML uncertainties.

### 4.2.2. Software Reliability Methods

In this section, several available software reliability methods are reviewed with the objective to classify potential techniques that can be used to quantify software failure probabilities of digital I&C systems at NPPs. Generally, there are two types of digital I&C systems at an NPP: a non-safety I&C system, such as the feedwater control system, and a safety system, such as the reactor trip system (RTS). Compared to non-safety systems, safety systems may have different failure modes. For instance, the RTS may fail to provide a trip signal when needed or send a spurious trip signal when not needed. Given the occurrence of an initiating event that leads to the need for a reactor trip, the first failure mode should be modeled in the PRA model in terms of a demand failure probability. And the latter failure mode should be identified as one of the initiating events, just as a failure of the feedwater system is modeled. This example shows that even for a single digital system, a different reliability method may be needed for application to different failure modes of interest [75].

Currently, none of existing methods becomes a general agreement for software reliability analysis of NPP digital I&C systems. Three types of these methods have been reviewed in [75]:

1. Reliability Growth Methods [76]. These are time-based methods that estimate software failure rates using test data. They can also predict the time to next failure and the required time to remove all faults. Reliability growth models are based on the sequence of times between observed and repaired failures. Different reliability growth methods contain different assumptions about how variables, such as failure rate and expected number of failures, change with time. The obtained failure rates from the reliability growth methods are usually in good agreement with the actual data. However, because the method is driven by test-failure data, the reliability of the software is not very high considering the significant testing impact. Therefore, for nuclear applications requiring high reliability, these models become unfeasible. Meanwhile, the testing environment may not fully represent the actual operational condition. Where the software is exposed during operation, the estimated failure rates may not accurately reflect the real number.

2. Bayesian Belief Network methods [77]. These are methods that use a probabilistic graphic model to describe a set of random variables and their conditional independencies using a directed acyclic graph. As a general method, BNN can model subjective opinions of experts and combine disparate information. It can also account for uncertainties in parameters within the scope of the model [78]. Moreover, instead of having full dependency among the random variables, BNN has relatively simple connections with localized conditional probability distributions, but the expert elicitation can be difficult to quantify when determining these discrepancies. Moreover, due to a lack of explainability of DDMs and ANN-based DTs, the collection and quantification of evidence can be very challenging.

3. Test-Based Methods [79]. Standard statistical methods are employed to run several tests and get the number of measured failures. However, similar to the problem seen in the reliability growth method, the operational condition may not be well defined, resulting to the inaccurate testing results generated. Moreover, because NPPs usually require high reliability for safety-critical software with very small failure probability(e.g., $10^{-5}$) [80], to gain statistical confidence in such a parameter, at least $3 \times 10^5$ successful tests need to be conducted to reach a confidence level of 0.95.

Although there is no agreement yet on which method to use, it has been agreed that software failure could and should be analyzed probabilistically. Some desirable characteristics are suggested in [75]: the method should be able to demonstrate high reliability of a safety-critical system and the method should account for the differences between testing conditions and operational profiles. A similar conclusion was made in the Workshop on Philosophical Basis for Incorporating Software Failures in a PRA [81]. The panelists universally agreed that: (1) software fails; (2) the occurrence of software failures can be treated probabilistically; (3) it is meaningful to use software failure rates and probabilities; and (4) software failure rates and probabilities can be included in reliability models of digital systems. According to the reviews in [82], software CCFs have been modeled in NPP PRA because they lead to the top event directly. Software CCF is generally modeled between processors with redundant designs and functions with the same application software and platform. This is similar to the module level of hardware failure depending on how many details are required to describe the features affecting the reliability of a software-based system. In [83], four different software failures were considered in the design-phase PRA conducted for an automation renewal of the Loviisa NPP: (1) independent failure; (2) CCF of a single automation system; (3) CCF of 16 programmed systems with the same platforms and/or software; and (4) CCF of programmed systems with different platforms and/or software.

### 4.2.3. Integrated Risk Assessment Strategy for Digital I&C Systems

In 2019, the Risk-Informed Systems Analysis (RISA) Pathway of the U.S. Department of Energy's (DOE's) Light Water Reactor Sustainability (LWRS) Program launched a project that aims to provide solid technical supports and risk-informed insights for the development and deployment of safe, secure, efficient and licensable digital I&C technologies in NPPs. [84] [85] An integrated Risk Assessment process for Digital I&C (RADIC) was proposed in this project including two phases: risk analysis and risk evaluation. Risk analysis aims to identify hazards of digital-based SSCs, estimate their failure probabilities, and analyze relevant consequences by performing hazard analysis, reliability analysis, and consequence analysis. The results from the risk analysis are compared with the specific risk-acceptance criteria in the risk evaluation. The schematic of the risk-assessment strategy for digital I&C systems is displayed in Figure 8. The tasks of the RADIC process are to evaluate whether the risk from digital failures can be accepted at the individual, system, and plant levels.

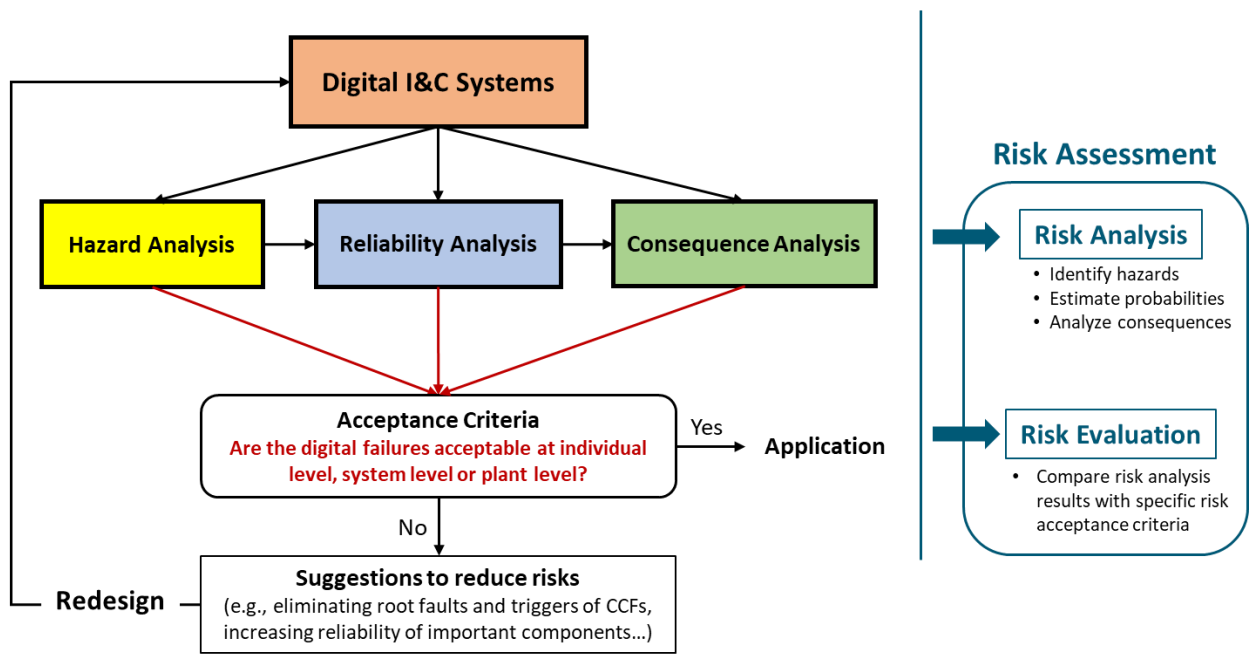

Figure 8. Schematic of RADIC process.

As the first part of risk analysis, hazard analysis needs to identify all potential software failures and hardware failures. Acceptance criterion is defined as, "Does the individual digital failure directly lead to the loss of function of the digital system?" The key outcome of hazard analysis is the integrated fault tree that includes both hardware and software failures, and its cut set for the top event. The top event is normally set as the loss of function of the target digital system. The "individual digital failure" in acceptance criterion refers to every basic event in this integrated fault tree for this specific top event, which can be a hardware failure, software failure, or human error. If the occurrence of this individual failure (e.g., software CCF) can result in the top event regardless of the occurrences of other basic events, its risk is not acceptable for the acceptance criterion. Accordingly, a redundancy-guided systems theoretic approach was developed safety-related digital I&C systems for supporting I&C designers and engineers to address both hardware and software CCFs, and to qualitatively analyze their effects on system availability. It also provides a technical basis for implementing following reliability and consequence analyses of unexpected software failures and recommending on the optimization of D3 analyses in a cost-efficient way. Targeting at the complexity of redundant designs in safety-related digital I&C systems, this approach integrates STPA, FTA and HAZCADS to identify software CCFs effectively by reframing STPA in a redundancy-guided way: (1) framing the complexity of the redundancy problem in a detailed representation; (2) clarifying the redundancy level using FTA before applying STPA; (3) building a redundancy-guided multilayer control structure; and (4) locating software CCFs for different levels of redundancy. This approach has been demonstrated and applied for the hazard analysis of a four-division digital RTS [86] and a four-division, digital, engineered safety features-actuation system [87].

The second part in risk analysis is reliability analysis with the tasks of: (1) quantifying the probabilities of basic events of the integrated FT from the hazard analysis; (2) determining the optimal basic component combinations for prevention and mitigation; and (3) estimating the probabilities of the consequences of digital system failures. The respective acceptance criterion is defined as, "How reliable is the digital system with the identified digital failures existing?" According to [84], quantification of failure probabilities is the main outcome for both single failure events and consequences. For different designs and requirements from licensing regulators, the set point for reliability probability should integrate the efforts and experiences from industry, regulators, and researchers. Bayesian networks and CCF modeling methods are incorporated to estimate the failure probabilities of basic events, particularly the software CCFs.

As the third and final part, consequence analysis should be implemented to evaluate the impact of consequences of digital failures on plant responses. The respective acceptance criterion is defined as, "Are the consequences of individual digital failures acceptable at the plant level?" The main concern is that some software failures have the potential to initiate an unanalyzed event or scenario that may not be analyzed before and, therefore, to threat reactor safety, such as by core damage or a large early release. The PRA results from the previous reliability analyses are supposed to provide different risk-significant accident scenarios for the multi-physics best-estimate plus uncertainty analysis. The capability has been built by different platforms such as the INL-developed LOTUS [88] and RAVEN [18] [19] .

## 5    CONCLUSIONS

To achieve an alignment of nuclear power plant (NPP) safety design, evaluation, operator training, and emergency management, a NAMAC system is developed to furnish recommendations to operators for effective actions that will achieve particular goals, based on the NAMAC's knowledge of the current plant state, prediction of the future state transients, and to reflect the uncertainties that complicate the determination of mitigating strategies. Such knowledge is extracted from the knowledge base and stored in the DTs of various functions. Considering the capability of data-driven algorithms in capturing nonlinear relationships with the complex reactor system and supporting real-time operations, ML algorithms (e.g., ANNs have been used in the NAMAC system as the DTs for diagnosis and prognosis). To ensure the consistency and transparency for the development of DTs, a DAP is suggested to guide the development and assessment of DTs according to the target expectations as set out in the planning stage. In this process, a two-tiered approach, one based on the top-down or system approach, the other on the bottom-up or separate approach, is predicted to minimize arbitrariness in deriving acceptance criteria, to develop a methodology that is transparent, consistent, and practical, and to provide a technical justifiable basis for deciding the adequacy of DTs and NAMAC for the target applications in a timely and cost-effective manner. It is found that UQ methods play key roles in the bottom-up approach, while software risk analysis is critical to the top-down approach. To guide the selection and further development of specific techniques, this study reviews a class of UQ and software risk analysis.

It is found that classical UQ techniques, including Bayesian inference, interval analysis, and fuzzy theory can be used to support the bottom-up approach in the DT-DAP. However, their applicability depends on the requirements in computational costs, target task and system complexity, and the availability of prior knowledge. Moreover, classical UQ techniques usually do not consider data quality, data quantity, or domain knowledge, and thus have limited improvement to the generalization capability of ANNs. Meta-learning, especially the transfer-learning techniques, can be used to improve the learning and generalization capability of ANNs. It is believed that as more tasks are learned and more prior knowledge is stored by the meta learner, the new task will become more similar and the generalization capability of ANN-based DTs will be naturally improved.

During the operation of ML-based DTs and respective autonomous control systems that are designed for safety purposes, crucial software CCFs may occur in different DTs for different intended uses in a NAMAC system or in redundant DTs for the same intended use. The capability of an integrated risk-informed analysis, such as the RADIC process, is significantly needed to meet the regulatory and industrial requirements for licensing and deployment of intelligent digital I&C systems (e.g., NAMAC system).

A systematic hazard analysis should be applied to construct a knowledge base of plant anomaly causes, including all potential failure modes (e.g., digital and mechanical malfunctions, human errors, physical and cyberattacks, environmental disturbance, etc.). Respective probabilities of these root causes should be provided by reliability analysis to evaluate the reliability of DT and the NAMAC system. Considering that some failures, especially software CCFs in DTs, may introduce some unanalyzed initiating events to the autonomous control systems, results from consequence analysis should be used to evaluate the impacts of these unanalyzed events and sequences on entire-plant responses. All the information is useful for the real-time operation of ML-based DTs.

# 6 REFERENCES

[1] L. Lin, P. Athe, P. Rouxelin, R. Youngblood, A. Gupta, J. Lane, M. Avramova and N. Dinh, "Development and assessment of a nearly autonomous management and control system for advanced reactors," *Annals of Nuclear Energy,* 2020.

[2] S. Cetiner, M. Muhlheim, A. Guler-Yigitoglu, R. Belles, S. Greenwood and T. Harrison, "Supervisory control system for multi-modular advanced reactors," Oak Ridge National Laboratory, Oak Ridge, 2016.

[3] K. M. Groth, M. R. Denman, J. N. Cardoni and T. A. Wheeler, "" Smart Procedures": Using dynamic PRA to develop dynamic context-specific severe accident management guidelines (SAMGs)," Sandia National Lab. (No. SAND2014-2207C), Albuquerque, 2014.

[4] M. Na, B. Upadhayaya, X. Xu and I. Hwang, "Design of a model predictive power controller for an SP-100 space reactor," *Nuclear Science and Engineering,* vol. 154, pp. 353-366, 2006.

[5] D. Lee, P. H. Seong and J. Kim, "Autonomous operation algorithm for safety systems of nuclear power plants by using long-short term memory and function-based hierarchical framework," *Annals of Nuclear Energy,* vol. 119, pp. 287-299, 2018.

[6] L. Lin, P. Rouxelin, P. Athe, J. Lane and N. Dinh, "Development and Assessment of Data-Driven Digital Twins in a Nearly Autonomous Management and Control System for Advanced Reactors," in *Procedings of the 2020 28th Conference on Nuclear Engineering*, Anaheim, California, 2020.

[7] K. Velten, Mathematical Modeling and Simulation: Introduction for Scientists and Engineers, Wiley, 2009.

[8] C. Chang and N. Dinh, "Classification of machine learning frameworks for data-driven thermal fluid models," *International Journal of Thermal Sciences,* vol. 135, pp. 559-579, 2019.

[9] B. Hanna, T. Son and N. Dinh, "An artificial intelligence-guided decision support system for the nuclear power plant management," in *18th International Topical Meeting on Nuclear Reactor Thermal Hydraulics*, Portland, WA, 2019.

[10] L. Lin, P. Athe, P. Rouxelin, J. Lane and N. Dinh, "Development and Assessment of a Nearly Autonomous Management and Control System during a Single Loss of Flow Accident," in *Proceedings of the 2020 28th Conference on Nuclear Engineering*, Anaheim, California, 2020.

[11] I. Goodfellow, Y. Bengio and A. Courville, Deep learning (adaptive computation and machine learning series), Cambridge: The MIT Press, 2016.

[12] D. Wolpert and W. Marcready, "No free lunch theorems for optimization," *IEEE Transactions on Evolutionary Computation,* vol. 1, no. 1, pp. 67-82, 1997.

[13] B. Boyack, I. Catton, R. Duffey, P. Griffth, K. Katsma, G. Lellouche, S. Levy, U. Rohatgi, G. Wilson, W. Wulff and N. Zuber, "Quantifying reactor safety margins part 1: An overview of the code scaling, applicability and uncertainty evaluation methodology," *Nuclear Engineering and Design,* vol. 119, no. 1, pp. 1-15, 1990.

[14] J. Friedman, T. Hastic and R. Tibshirani, The elements of statistical learning, New York: Springer, 2001.

[15] U.S. NRC, "Transient and accident analysis methods," U.S. Nuclear Regulatory Commission, Washington D.C., 2005.

[16] M. Li and I. Bolotnov, "The evaporation and condensation model with interface tracking," *International Journal of Heat and Mass Transfer,* vol. 150, p. 119256, 2020.

[17] J. Lane, J. Link, J. King, T. George and S. Claybrook, "Benchmark of GOTHIC to EBR-II SHRT-17 and SHRT45R tests," *Nuclear Technology,* 2020.

[18] C. Rabiti, A. Alfonsi, J. Cogliati, D. Mandelli, R. Kinoshita, S. Seb, C. Wang, P. W. Talbot and D. P. Maljovec, "RAVEN User Manual," INL, Idaho Falls, 2015.

[19] A. Alfonsi, C. Rabiti, D. Mandelli, J. Cogliati, C. Wang, P. Talbot, D. Maljovec and C. Smith, "RAVEN Theory Manual," INL, Idaho Falls, 2016.

[20] C. Sammut and G. I. Webb, Encyclopedia of machine learning, Boston, MA: Springer, 2010.

[21] P. Walley, Statistical Reasoning with Imprecise Probabilities, Chapman and Hall/CRC, 1990.

[22] C. Roy and W. Oberkampf, "A comprehensive framework for verification, validation, and uncertainty quantification in scientific computing," *Computer Methods in Applied Mechanics and Engineering,* vol. 200, pp. 2131-2144, 2011.

[23] W. Oberkampf and C. Roy, Verification and Validation in Scientific Computing, Cambridge University Press, 2010.

[24] P. Roache, Verification and Validation in Computational Science and Engineering, Albuquerque: Hermosa Publishers, 1998.

[25] R. Bank, "Hierarchical bases and the finite element method," *Acta Numerica,* vol. 5, pp. 1-43, 1996.

[26] P. Cavallo and N. Sinha, "Error quantification for computational aerodynamics using an error transport equation," *Journal of Aircraft,* vol. 44, pp. 1954-1963, 2007.

[27] I. Celik, U. Ghia, P. Roache and C. J. Freitas, "Procedure for estimation and reporting of uncertainty due to discretization in CFD applications," *Journal of Fluid Engineering*.

[28] P. Roache, K. Ghia and F. White, "Editorial policy statement on control of numerical accuracy," *Journal of Fluid Engineering,* vol. 108, p. 1, 1986.

[29] S. Ferson and L. Ginzburg, "Different methods are needed to propagate ignorance and variability," *Reliability Engineering and System Safety,* vol. 54, pp. 133-144, 1996.

[30] D. Sivia and J. Skilling, Data analysis: A Bayesian tutorial, Oxford, UK: Oxford University Press, 2006.

[31] Y. Liu, X. Sun and N. Dinh, "Validation and uncertainty quantification of multiphase-CFD solvers: A data-driven Bayesian framework supported by high-resolution experiments," *Nuclear Engineering and Design,* vol. 354, 2019.

[32] D. MacKay, "A Practical Bayesian Framework for Backpropagation Networks," *Neural Computation,* vol. 4, no. 3, pp. 448-472, 1992.

[33] R. M. Neal, "Bayesian Learning for Neural Networks," University of Toronto, Toronto, 1995.

[34] R. Zhao and Q. JI, "An Empirical Evaluation of Bayesian Inference Methods for Bayesian Neural Networks," in *NIPS Bayesian Deep Learning (BDL) Workshop*, 2018.

[35] C. Blundell, J. Cornebise, K. Kavukcuoglu and D. Wierstra, "Weight Uncertainty in Neural Networks," in *the 32nd International Conference on Machine Learning*, Lille, France, 2015.

[36] C. Louizos and M. Welling, "Multiplicative Normalizing Flows for Variational Bayesian Neural Networks," in *Proceedings of the 34th International Conference on Machine Learning*, 2017.

[37] N. Pawlowski, M. Rajchl and B. Glocker, "Implicit Weight Uncertainty in Neural Networks," in *Second workshop on Bayesian Deep Learning*, Long Beach, CA, 2017.

[38] J. Yao, W. Pan, S. Ghosh and F. Doshi-Velez, "Quality of Uncertainty Quantification for Bayesian Neural Network Inference," in *36th International Conference on Machine Learning*, Long Beach, CA, 2019.

[39] R. Muhanna, H. Zhang and R. Mullen, "Interval Finite Elements as a Basis for Generalized Models of Uncertainty in Engineering Mechanics," *Raliable Computing,* vol. 13, no. 2, pp. 173-194, 2007.

[40] R. Muhanna and R. Mullen, "Uncertainty in Mechanics Problems: Interval-based Approach," *Journal*, vol. 127, no. 6, pp. 557-566, Journal of Engineering Mechanics.

[41] H. Li, H. Li and Y. Du, "A Global Optimization Algorithm based on Novel Interval Analysis for Training Neural Networks," *Advances in Computation and Intelligence,* vol. 4683, pp. 286-295, 2007.

[42] E. d. Weerdt, Q. Chu and J. Mulder, "Neural Network Output Optimization Using Interval Analysis," *IEE Transactions on Neural Networks,* vol. 20, no. 4, pp. 638-653, 2009.

[43] H. Bandemer and S. Gottwald, Fuzzy sets, fuzzy logic, fuzzy methods with applications, Wiley, 1995.

[44] D. Dubois and H. Prade, "The three semantics of fuzzy sets," *Fuzzy Sets and Systems,* vol. 90, no. 2, pp. 141-150, 1997.

[45] R. Viertl, Statistical methods for non-precise data, CRC Press, 1996.

[46] J. Abonyi, Fuzzy Model Identification for Control, Boston, MA: Birkhauser, 2003.

[47] E. Lughofer, Evolving Fuzzy Systems – Methodologies, Advanced Concepts and Applications, Springer-Verlag Berlin Heidelberg, 2011.

[48] Z. Pezeshki and S. Mazinani, "Comparison of artificial neural networks, fuzzy logic and neuro fuzzy for predicting optimization of building thermal consumption: a survey," *Artificial Intelligence Review,* vol. 52, pp. 495-525, 2019.

[49] B. Lake, T. Ullman, J. Tenenbaum and S. Gershman, "Building machines that learn and think like people," *Behavior and Brain Science,* vol. 40, 2017.

[50] P. Brazdil, C. Giraud-Carrier, C. Soares and R. Vilalta, Metalearning: Applications to Data Mining, Berlin, Heidelberg: Springer-Verlag, 2009.

[51] J. Vanschoren, "Meta-Learning," in *Automated Machine Learning. The Springer Series on Challenges in Machine Learning*, Cham, Springer, 2019, pp. 35-61.

[52] F. Hutter, H. H. Hoos and K. Leyton-Brown, "Sequential Model-Based Optimization for General Algorithm Configuration," in *International Conference on Learning and Intelligent Optimization*, 2011.

[53] D. Jones, "A taxonomy of global optimization methods based on response surfaces," *Journal of Global Optimization,* vol. 21, pp. 345-383, 2001.

[54] J. Villemonteix, E. Vazquez and E. Walter, "An informational approach to the global optimization of expensive-to-evaluate functions," *Journal of Global Optimization,* 2006.

[55] N. Srinivas, A. Krause, S. Kakade and M. Seeger, "Gaussian process optimization in the bandit setting: no regret and experimental design," in *International Conference of Machine Learning*, 2010.

[56] C. Castiello, G. Castellano and A. Fanelli, "Meta-data: Characterization of input features for meta-learning," in *International Conference on Modeling Decisions for Artificial Intelligence (MDAI)*, 2005.

[57] Q. Sun and B. Pfahringer, "Pairwise meta-rules for better meta-learning-based algorithm ranking," *Machine Learning,* vol. 93, no. 1, pp. 141-161, 2013.

[58] V. Perrone, R. Jenatton, M. Seeger and C. Archambeau, "Multiple adaptive Bayesian linear regression for scalable Bayesian optimization with warm start," in *Conference on Neural Information Processing Systems*, 2018.

[59] H. Bao, N. Dinh, L. Lin, R. Youngblood, J. Lane and H. Zhang, "Using Deep Learning to Explore Local Physical Similarity for Global-scale Bridging in Thermal-hydraulic Simulation," *Annals of Nuclear Energy,* 2020.

[60] H. Bao, J. Feng, N. Dinh and H. Zhang, "Deep Learning Interfacial Momentum Closures in Coarse-Mesh CFD Two-Phase Flow Simulation Using Validation Data," *International Journal of Multiphase Flow,* p. 103489, 2020.

[61] S. Thrun and L. Pratt, "Learning to learn: Introduction and overview," in *Learning to learn*, Boston, MA, Springer, 1998, pp. 3-17.

[62] J. Schmidhuber, "Learning to control fast-weight memories: an alternative to dynamic recurrent networks," *Neural Computing,* vol. 4, no. 1, pp. 131-139, 1992.

[63] S. Bengio, Y. Bengio and J. Cloutier, "On the search for new learning rules for ann," *Neural Processing Letters,* vol. 2, no. 4, pp. 26-30, 1995.

[64] T. Runarsson and M. Jonson, "Evolution and design of distributed learning rules," in *IEEE Symposium on Combinations of Evolutionary Computation and Neural Networks*, 2000.

[65] A. Younger, S. Hochreiter and P. Conwell, "Meta-learning with backpropagation," in *International Joint Conference on Neural Networks*, 2001.

[66] Y. Ioannou, "Structural priors in deep neural networks," University of Cambridge, Cambridge, UK, 2017.

[67] S. Seo and Y. Liu, "Differentiable physics-informed graph networks," in *The international Conference on Learning Representations*, New Orleans, 2019.

[68] M. Raissi, A. Yazdani and G. Karniadakis, "Hidden fluid mechanics: learning velocity and pressure fields from flow visualizations," *Science,* vol. 367, no. 6481, pp. 1026-1030, 2020.

[69] S. A. Arndt and A. Kuritzky, "Lessons Learned from the U.S. Nuclear Regulatory Commission's Digital System Risk Research," *Nuclear Technology,* vol. 173, no. 1, pp. 2-7, 2010.

[70] U.S.NRC, "Research Information Letter 1002: Identification of Failure Modes in Digital Safety Systems – Expert Clinic Findings, Part 2," U.S.NRC, Washington, D.C..

[71] N. G. Leveson and J. P. Thomas, STPA Handbook, March 2018.

[72] J. Thomas, F. L. d. Lemos and N. Leveson, "Evaluating the Safety of Digital Instrumentation and Control Systems in Nuclear Power Plants," MIT, Cambridge, Massachusetts, 2012.

[73] N. Leveson, C. Wilkinson, C. Flemming, J. Thomas and I. Tracy, "A Comparison of STPA and the ARP 4761 Safety Assessment Process," Massachusetts Institute of Technology, Cambridge, MA, 2014.

[74] A. Williams, A. Clark, A. Muna and M. Gibson, "Hazard and Consequence Analysis for Digital Systems – A New Approach to Risk Analysis in the Digital Era for Nuclear Power Plants," in *2018 ANS Winter Meeting and Nuclear Technology Expo*, Orlando, Florida, November 11-15, 2018.

[75] T.-L. Chu, J. Lehner and M. Yue, "Review of Quantitative Software Reliability Methods," Brookhaven National Laboratory, Long Island, NY, 2011.

[76] H. Son, H. Kang and S. Change, "Procedure for Application of Software Reliability Growth Models to NPP PSA," *Nuclear Engineering and Technology,* vol. 41, no. 8, 2009.

[77] H.-S. Eom, G. Park, H. Kang and S. Jang, "Reliability Assessment Of A Safety-Critical Software By Using Generalized Bayesian Nets," in *6th ANS Topical Meeting on Nuclear Plant Instrumentation, Controls and Human-Machine Interfaces Technologies (NPIC&HMIT 2009)*, Knoxville, Tennessee, 2009.

[78] J. Pearl, Bayesian networks: A model of self-activated memory for evidential reasoning, Los Angeles: University of California, 1985.

[79] B. G. Dahll and U. Pulkkinen, "Software-based system reliability," Nuclear Energy Agency, 2007.

[80] EPRI, "Estimating failure rates in highly reliable digital systems," EPRI, Palo Alto, CA, 2010.

[81] T. Chu, G. Martinez-guridi, M. Yue, P. Samanta, G. Vinod and J. Lehner, "Workshop on Philosophical Basis for Incorporating Software Failures in a Probabilistic Risk Assessment," Brookhaven National Laboratory, 2009.

[82] O. Backstrom, J. Holmberg, M. Jockenhovel-Barttfeld, M. Porthin, A. Taurines and T. Tyrvainen, "Software reliability analysis for PSA: failure mode and data analysi," Nordic Nuclear Safety Research (NSK), Roskilde, Denmark, 2015.

[83] K. Jänkälä, "Reliability of New Plant Automation of Loviisa NPP," in *Proceedings of the DIGREL seminar "Development of best practice guidelines on failure modes taxonomy for reliability assessment of digital I&C systems for PSA*, Espoo, Finland, 2011.

[84] H. Bao, H. Zhang and K. Thomas, "An Integrated Risk Assessment Process for Digital Instrumentation and Control Upgrades of Nuclear Power Plants," Idaho National Laboratory, Idaho Falls, ID, 2019.

[85] H. Bao, T. Shorthill and H. Zhang, "Redundancy-guided System-theoretic Hazard and Reliability Analysis of Safety-related Digital Instrumentation and Control Systems in Nuclear Power Plants," Idaho National Laboratory, Idaho Falls, 2020.

[86] T. Shorthill, H. Bao, H. Zhang and H. Ban, "A Redundancy-Guided Approach for the Hazard Analysis of Digital Instrumentation and Control Systems in Advanced Nuclear Power Plants," *arXiv.org,* 2020.

[87] H. Bao, T. Shorthill and H. Zhang, "Hazard Analysis for Identifying Common Cause Failures of Digital Safety Systems using a Redundancy-Guided Systems-Theoretic Approach," *Annals of Nuclear Energy,,* vol. 148, p. 107686, 1 December 2020.

[88] H. Zhang, R. Szilard, S. Hess and R. Sugrue, "A Strategic Approach to Employ Risk-Informed Methods to Enable Margin Recovery of Nuclear Power Plants Operating Margins," Idaho National Laboratory, Idaho Fall,s ID, September 2018.